# Joint analysis of structural connectivity and cortical surface features: correlates with mild traumatic brain injury

Cailey I. Kerley[1], Leon Y. Cai[2], Chang Yu[3], Logan M. Crawford[4], Jason M. Elenberger[4], Eden S. Singh[4], Kurt G. Schilling[5], Katherine S. Aboud[6], Bennett A. Landman[1,2,3,5,6], and Tonia S. Rex[4]

[1]Department of Electrical Engineering, Vanderbilt University; [2]Department of Biomedical Engineering, Vanderbilt University; [3]Department of Computer Science, Vanderbilt University; [4]Department of Ophthalmology and Visual Sciences, Vanderbilt University Medical Center; [5]Vanderbilt University Institute of Imaging Science, Vanderbilt University; [6]Vanderbilt Brain Institute, Vanderbilt University

## ABSTRACT

Mild traumatic brain injury (mTBI) is a complex syndrome that affects up to 600 per 100,000 individuals, with a particular concentration among military personnel. About half of all mTBI patients experience a diverse array of chronic symptoms which persist long after the acute injury. Hence, there is an urgent need for better understanding of the white matter and gray matter pathologies associated with mTBI to map which specific brain systems are impacted and identify courses of intervention. Previous works have linked mTBI to disruptions in white matter pathways and cortical surface abnormalities. Herein, we examine these hypothesized links in an exploratory study of joint structural connectivity and cortical surface changes associated with mTBI and its chronic symptoms. Briefly, we consider a cohort of 12 mTBI and 26 control subjects. A set of 588 cortical surface metrics and 4,753 structural connectivity metrics were extracted from cortical surface regions and diffusion weighted magnetic resonance imaging in each subject. Principal component analysis (PCA) was used to reduce the dimensionality of each metric set. We then applied independent component analysis (ICA) both to each PCA space individually and together in a joint ICA approach. We identified a stable independent component across the connectivity-only and joint ICAs which presented significant group differences in subject loadings (p<0.05, corrected). Additionally, we found that two mTBI symptoms, slowed thinking and forgetfulness, were significantly correlated (p<0.05, corrected) with mTBI subject loadings in a surface-only ICA. These surface-only loadings captured an increase in bilateral cortical thickness.

**Keywords:** mild traumatic brain injury, independent component analysis, connectome, cortical surface, MRI

## 1. INTRODUCTION

Mild traumatic brain injury (mTBI) is a disruption of normal brain function caused by any injury to the head. It is estimated that in North America, mTBI has an incidence rate of more than 600 per 100,000 inhabitants [1]. Military personnel and veterans are particularly burdened by this disorder; 11%-23% of soldiers are expected to experience a traumatic brain injury while deployed, with the majority of cases classified as mild [2]. Additionally, approximately half of all mTBI patients are left with chronic symptoms that persist long after the injury's acute phase [3]. Due to the prevalence of mTBI and this long-term adverse symptomology, there is an urgent need for advanced imaging methods that can localize mTBI effects in the brain.

mTBI has previously been linked to abnormalities of the cortical surface [4] and disrupted white matter pathways [5]. One hypothesis is that mTBI may be classified as a "disconnection syndrome" due to the prevalence of these white matter disconnections [6]. In a previous work, we demonstrated the potential of combining cortical surface and sensory white matter pathway features in a discriminatory model of mTBI [7]. In this article, we build on that work by investigating a joint analysis of magnetic resonance imaging (MRI) derived cortical surface and structural connectivity features in mTBI and control subjects. This exploratory analysis aims to identify joint cortical surface and structural connectivity changes associated with mTBI and related symptoms, which may be tested in a larger sample.

Briefly, this work involved first extracting a) a set of 588 cortical shape metrics and b) a set 4,753 structural connectivity metrics from 98 cortical surface regions for each subject. The dimensionality of each metric set was reduced via principal component analysis (PCA). Independent component analysis (ICA) was then applied to the PCA spaces of the cortical surface metrics, structural connectivity metrics, and a joint metric set (a concatenation of cortical surface and structural connectivity PCA spaces). Subjects' independent component (IC) loadings were compared across the mTBI and control groups to assess group differences. The correlation between mTBI subjects' IC loadings and self-reported symptom severity scores was also examined.

## 2. METHODS

### 2.1 Imaging and symptom data

This exploratory study considered a cohort of 38 subjects, of whom 12 were mTBI subjects (previous mTBI diagnosis confirmed via Electronic Medical Record with a Glasgow Coma Score in the range 13-15) and 26 were controls with no history of mTBI or audiovisual problems. All subjects had both T1-weighted MRI (T1w) and diffusion weighted MRI (DWI) scans acquired during a single session at the Vanderbilt University Institute of Imaging Science. Three DWI volumes were acquired for each subject: a 32-direction b=1000 s/mm$^2$, a 64-direction b=2000 s/mm$^2$, and a corresponding b=0 s/mm$^2$ volume. These 32 and 64 shell DWI volumes were concatenated and corrected for eddy current distortions and patient movement following the protocol in [8].

For all mTBI subjects, chronic mTBI symptoms were assessed via the Neurobehavioral Symptom Inventory [9], a questionnaire that tracks 22 TBI symptoms: dizziness, loss of balance, poor coordination, headaches, nausea, vision problems, light sensitivity, hearing difficulty, noise sensitivity, numbness, taste or smell changes, appetite changes, poor concentration, forgetfulness, difficulty making decisions, slowed thinking, fatigue, difficulty sleeping, anxiety, feeling depressed, irritability, and frustration. The severity of each symptom was ranked by the subject on a scale of 1 to 5, where 1 was unaffected and 5 was a significant impact on daily life.

### 2.2 Metric generation

The cortical surface and structural connectivity imaging metric generation processes are shown in Figure 1. First, the T1w volume was segmented into 132 BrainColor regions via multi-atlas segmentation [10], and MaCRUISE [11] was used to reconstruct cortical surfaces from the volumetric segmentation. To define boundaries of regions of interest (ROI) on the cortical surfaces, the cortical surfaces were mapped to a unit sphere and then rigidly aligned to the adult template created by [12]. For each individual cortical surface, the cortical ROIs were determined via spherical convolutional networks [13]. The resulting labels are a subset of BrainColor, which contains 98 regions out of the original 132. A cortical surface analysis was then performed, yielding six shape metrics averaged over each cortical surface region, including mean curvature, shape index [14], sulcal depth [15], cortical thickness [11], shape complexity index [16], and local gyrification index [17], [18]. This resulted in a total of 588 surface metrics.

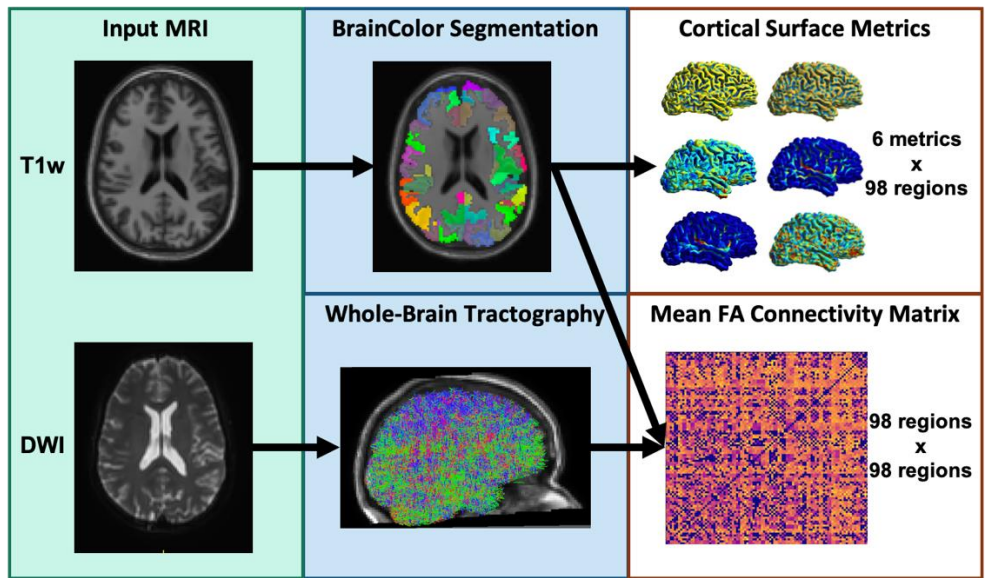

Figure 1. Surface and connectivity metric generation. The T1w volume is segmented into 132 BrainColor regions, out of which 98 cortical surface regions are kept. A cortical shape analysis is performed on the 98 regions, yielding 6 shape metrics per region: mean curvature, shape index, sulcal depth, cortical thickness, shape complexity index, and local gyrification index. Whole brain tractography is performed on the DWI volume. This tractogram is used to construct a connectivity matrix for the 98 surface regions, where connection strength is equivalent to mean fractional anisotropy (FA) along streamlines connecting each pair of regions.

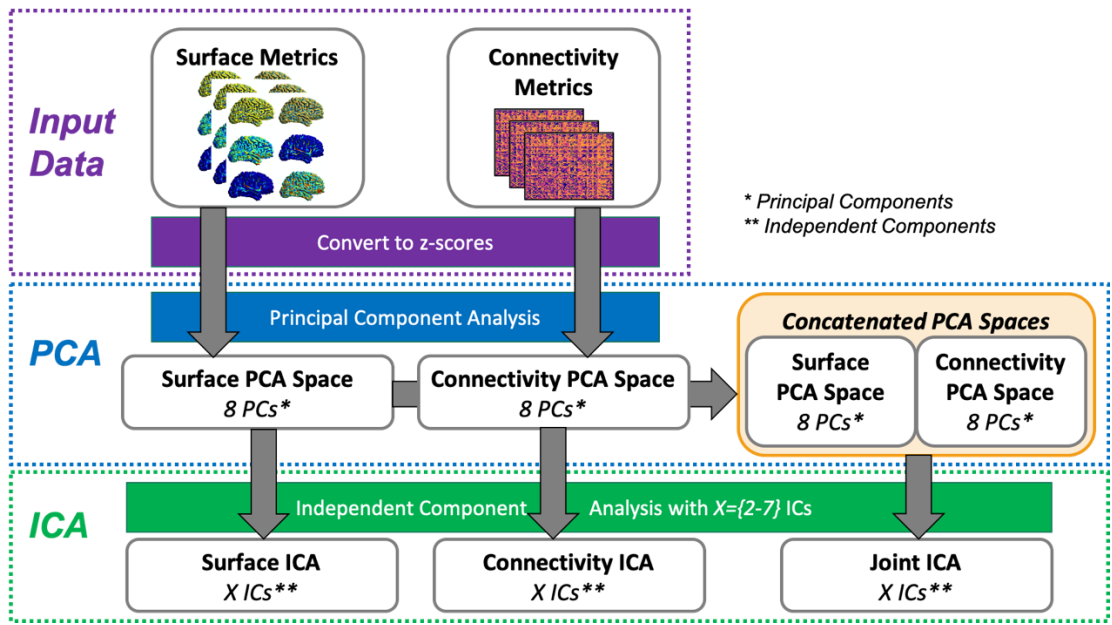

Figure 2. Metric analysis pipeline. All metrics are first normalized by converting the raw data to z-scores. Principal component analysis (PCA) is then performed separately on the surface and connectivity metrics; the dimensionality of both metric sets is reduced to the 8 most principal components from their respective PCA spaces. These two PCA spaces are further reduced via independent component analysis (ICA) to $X$ independent components (ICs). The value of X is swept across all possible values, from 2 to 7 ICs. To perform a joint analysis of the surface and connectivity metrics, their 8-component PCA spaces are concatenated, and ICA is again performed on this joint data set for X={2-7} ICs.

To derive the structural connectivity metrics, each subject's 98-region BrainColor segmentation was registered to their DWI volume, so that the same cortical surface regions were used for both the surface and connectivity metrics. Next, a whole-brain tractogram was generated for each subject. The response function for tracking was estimated via the iterative Tournier algorithm [19], after which the fiber orientation distribution was calculated using spherical deconvolution. The iFOD2 algorithm [20] was then used to perform probabilistic tractography and construct a one million streamline whole-brain tractogram seeded at random within a whole-brain mask. Finally, a connectivity matrix was generated for each subject, where the nodes were the 98 cortical surface regions, and the edges were defined as mean fractional anisotropy (FA) along the streamlines connecting each pair of regions. This mean FA linking pairs of regions was used as our structural connectivity metric. Self-connections were excluded, and streamline direction was ignored, yielding 4,753 metrics. All tractography and connectivity matrix operations were performed via the MRTrix3 package [21].

**2.3 Metric analysis**

An overview of the imaging metric analysis is shown in Figure 2. Briefly, this analysis pipeline includes metric normalization and dimensionality reduction via principal component analysis (PCA) as preprocessing steps before applying independent component analysis (ICA). ICA can be sensitive to high variance, so PCA was applied prior to the ICA step in order to reduce variance in the dataset. Additionally, reconstruction ICA was used; this variant on traditional ICA optimizes on a soft reconstruction constraint [22]. Reconstruction ICA is a desirable method for this application, as it was designed to extract stable features which are not as sensitive to variance in the underlying data as traditional ICA. The following description includes more specific details of our method design.

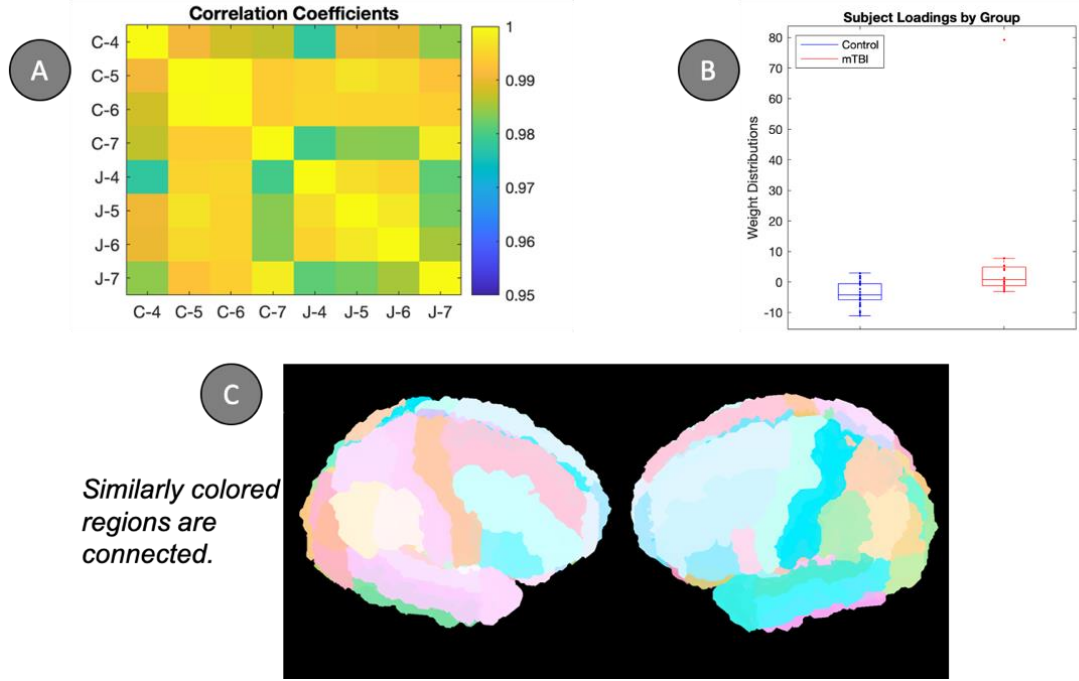

Figure 3. An independent component presenting significant group differences. A singular independent component was consistently found across both the connectivity-only and joint ICAs. A) Extremely high correlation coefficients are seen for this component across connectivity-only and joint ICAs with X = {4,5,6,7} ICs. (Rows/columns of this matrix are denoted by "ICA type – X", with C connectivity-only and J = joint; so "C-4" means connectivity-only ICA with X=4.) All correlations are statistically significant ($p <<< 0.001$, corrected). B) The subject loadings for this component present a statistically significant ($p < 0.05$, corrected) difference between the control and mTBI populations. C) The independent component back-projected into connectivity data space; the component is visualized on a representative T1w volume, where similarly colored regions are connected (for a full description of the visualization procedure, see section 2.4).

The 588 surface and 4,753 connectivity metrics were normalized via conversion to z-score representation; z-scores are calculated for each individual metric by subtracting the group's mean and dividing by the group's standard deviation. Next, the dimensionality of each metric set was reduced via PCA. This reduction revealed that principal components (PCs) beyond the top 8 in both the surface and connectivity PCA spaces represented negligible explained variance in the data. Based on this and the need to balance the metric sets in the joint analysis, the PCs were narrowed down to the top 8 for both the surface and connectivity PCA spaces. A joint PCA space was then created by concatenating the surface and connectivity PCs, yielding a 16-component joint space. Finally, the reconstruction ICA was applied to the three PCA spaces. To fully explore all possible ICA features, X, the number of independent components (ICs), was varied from 2 to 7 for all three PCA spaces. To differentiate between these, we will refer to each individual ICA in the form of "ICA type – X", where possible types are *S* (surface-only ICA), *C* (connectivity-only ICA), and *J* (joint ICA); for example, S-6 refers to surface-only ICA with 6 ICs. Note that 1 and 8 ICs were excluded because singular reduction (1 IC) and no reduction (8 ICs) were considered uninteresting edge cases.

Each ICA produces two important outputs: ICs and subject loadings. ICs are the set of independent features learned from the PCA space; they may be back-projected into the original data space to investigate relationships between brain regions. These ICs are related to study subjects according to each subject's set of IC loadings. These loadings are a set of weights (one per IC) that correspond to how strongly each IC is represented in an individual subject's input signal. So, while ICs describe overall patterns in the data, subject loadings describe variations of those patterns in individuals. The following sections explain how these two outputs are investigated.

### 2.4 Visualizing independent components

The many sets of ICs derived in section 2.3 would be largely useless without a means for interpretation. To facilitate visual interpretation, all ICs were back-projected into the original cortical surface and/or structural connectivity spaces and scaled to the range [0,1]. All ICs were visualized using the 98-region cortical surface BrainColor segmentation of a single representative T1w volume. Since the surface data space contained only six metrics per region, surface ICs were visualized on six individual three-dimensional surface renderings, one for each metric. In these renderings, region-wise IC coefficients were mapped to each region's color.

Due to the high dimensionality of the structural connectivity data space, the connectivity ICs required a more specialized visualization approach. In this visualization, the entire connectome was presented on a single T1w surface, with region color and transparency denoting region-wise and overall connectivity, respectively. To illustrate, consider a single IC back-projected to a connectivity matrix *C*. First, all self-connections are set to 1 (fully connected). *C* is then converted to a dissimilarity matrix *D* by subtracting all elements from 1. Nonmetric multidimensional scaling with Kruskal's normalized stress1 criterion [23] is applied to *D,* so that each region is mapped to a two-dimensional space *D'* in which the Euclidian distance between any two points approximates a monotonic transformation of their corresponding dissimilarity in *D*. The two dimensions of *D'* are then used as the a* (green-red) and b* (blue-yellow) dimensions in the CIELAB colorimetric system [24]. The last dimension, *L (lightness), is calculated by summing a region's edges in *C* as a percentage of the summed edges of the most connected region. Thus, *C* is converted to a region-wise CIELAB color map. A three-dimensional T1w cortical surface is then generated in which each region is colored according to this color map, and each region's transparency is scaled according to its L* dimension. In this way, the IC is visualized on a single surface, such that similarly colored regions are more connected to each other, and more opaque regions are more connected overall.

## 3. RESULTS

### 3.1 Group differences

A primary aim of this analysis was to identify cortical surface and structural connectivity changes associated with mTBI. To this end, we investigated group differences by comparing subject IC loadings between the mTBI and control groups for each IC within each ICA. This comparison was accomplished via a Wilcoxon rank-sum test, with Bonferroni multiple comparisons correction applied to all tests within each ICA (e.g. Bonferroni was applied for 6 comparisons in J-6, and for 7 comparisons in J-7). Through this method, C-4, C-5, C-6, C-7, J-4, J-5, J-6, and J-7 were all found to have an IC that presented a statistically significant group difference in subject loadings ($p < 0.01$, corrected). These 8 ICs were compared using a Pearson correlation and all were found to be highly correlated ($p \ll 0.001$, corrected), as shown in Figure 3A. The most correlated IC out of this group (C-5) was selected to demonstrate the significant group difference in subject loadings (Figure 3B) and to visualize this IC's connectome (Figure 3C).

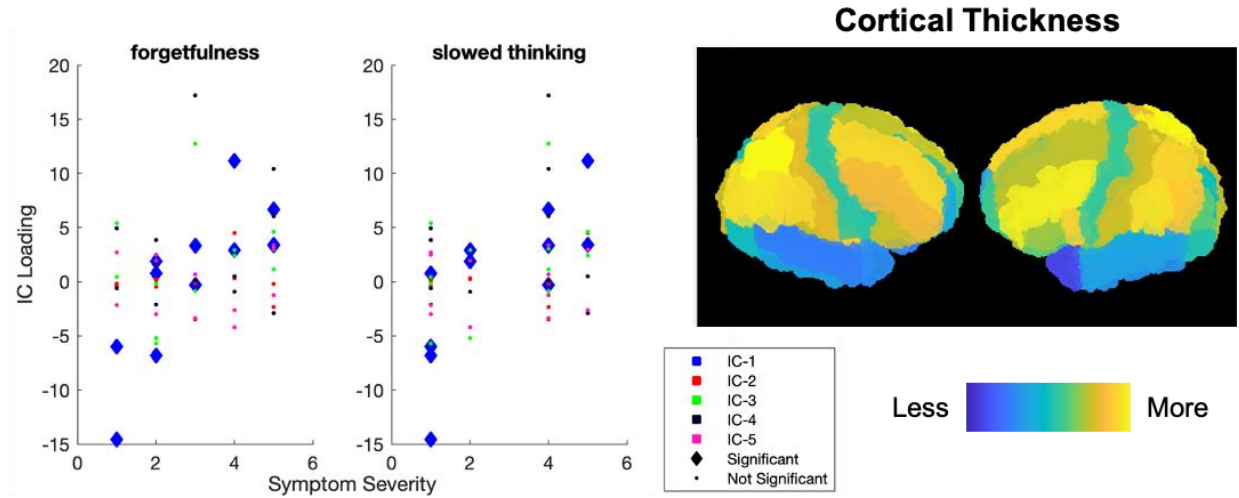

Figure 4. An independent component presenting significant symptom correlations. Across all ICAs performed, only a single IC was found to be significantly correlated to mTBI symptom scores. This IC is from the surface-only ICA performed with 5 ICs. The two plots on the left in this figure show the statistically significant ($p < 0.01$, corrected) positive correlations between mTBI subject IC loadings and symptom severity scores for forgetfulness and slowed thinking. The IC was back-projected into the cortical surface metrics dataspace. The cortical thickness metric is visualized here on a representative T1w volume; the magnitude of region IC coefficients are encoded via color, with blue corresponding to lower absolute thickness and yellow corresponding to increased absolute thickness.

The stability of this significant IC across the 8 ICAs suggests that the pattern of connectivity seen in Figure 3C is a legitimate feature, not an artifact of ICA's optimization scheme. Furthermore, the IC's subject loadings show that this connectivity pattern is more prominent in mTBI subjects than in controls, implying that the connections may be indicative of mTBI pathology. Table 1 presents the seven region-to-region connections with the largest weights in the IC from C-5. The strongest connection involves the right parietal operculum, which is believed to be involved with touch and pain perception [25], and the left occipital fusiform gyrus, a region involved in high level visual processing [26]. Several other regions in this list are associated with visual processing (right occipital fusiform gyrus [27], right calcarine cortex [28]), working memory (right/left superior frontal gyrus medial segment [29][30]), and other sensory functions (right angular gyrus [31], left opercular part of the inferior frontal gyrus [32]), all cognitive functions which have been known to be negatively impacted by mTBI [9].

Table 1. Top seven most highly weighted connections in an IC from C-5 associated with significant group differences

| Weight Rank | ROI 1 | ROI 2 |
|---|---|---|
| 1 | Right parietal operculum | Left occipital fusiform gyrus |
| 2 | Right occipital fusiform gyrus | Left superior frontal gyrus medial segment |
| 3 | Left lateral orbital gyrus | Right angular gyrus |
| 4 | Right posterior orbital gyrus | Right central operculum |
| 5 | Right calcarine cortex | Left anterior orbital gyrus |
| 6 | Right superior frontal gyrus medial segment | Left frontal operculum |
| 7 | Left opercular part of the inferior frontal gyrus | Right opercular part of the inferior frontal gyrus |

### 3.2 Symptom correlations

To further explore the relationship between these IC features and mTBI symptomology, Spearman rank correlations were calculated between mTBI subject symptom scores and IC loadings for all ICs across all ICAs. Out of all tests, a single IC in S-5 was found to be significantly correlated with two symptoms after Bonferroni correction: forgetfulness and slowed thinking ($p < 0.01$, corrected for 110 comparisons). Figure 4 shows these correlations between S-5 subject IC loadings and symptom scores for the two symptoms, along with the IC back-projected to the cortical thickness metric. This positive correlation between symptom severity scores and subject loadings indicates that the IC presents most in subjects with more severe forgetfulness and slowed thinking. Combining this with the IC's pattern cortical thickness pattern suggests that these mTBI symptoms may be related to increased bilateral cortical thickness. This finding is promising, since other studies have also found increased prefrontal cortical thickness to be associated with poorer outcomes in mTBI subjects one year after the acute injury [4].

## 4. DISCUSSION

We have presented a joint analysis of structural connectivity and cortical surface structure in mTBI via ICA. We demonstrated the potential of this framework for hypothesis generation regarding mTBI pathologies via two significant IC findings. First, our analysis identified a stable IC that presented significant differences in IC loadings between mTBI subjects and controls, in both connectivity-only and joint ICA. This finding revealed structural connectivity changes in individual cortical surface regions that are potentially linked to mTBI pathologies. Additionally, we found a strong correlation between mTBI IC loadings and symptom severity scores for the slowed thinking and forgetfulness symptoms; examining the surface metric representation of this IC suggests that the severity of these symptoms is linked to a relative increase in cortical thickness among mTBI subjects.

In summary, the proposed framework was found to be effective for detecting potential structural connectivity and cortical surface abnormalities in mTBI. A major advantage of this framework was the ability to back-project ICs into data space, allowing for anatomical interpretations of group differences and symptom correlations. This explanatory power makes ICA a desirable method for exploring disease pathologies compared to "black box" neural networks.

## ACKNOWLEDGEMENTS

Support for this work included funding from DoD grants W81XWH-15-1-0096, W81XWH-17-2-0055, NEI grants R01 EY022349, U24EY029893, P30EY008126, Retired Major General Stephen L. Jones, MD Fund, and Research Prevent Blindness, Inc. (VEI), as well as NSF CAREER 1452485 (Landman) and NIH grant T32EB001628 (Schilling). This study was in part using the resources of the Advanced Computing Center for Research and Education (ACCRE) at Vanderbilt University, Nashville, TN. This project was supported in part by ViSE/VICTR VR3029 and the National Center for Research Resources, Grant UL1 RR024975-01, and is now at the National Center for Advancing Translational Sciences, Grant 2 UL1 TR000445-06.

## REFERENCES

[1] J. D. Cassidy et al., "Incidence, risk factors and prevention of mild traumatic brain injury: results of the WHO collaborating centre task force on mild traumatic brain injury," *J. Rehabil. Med.*, vol. 36, pp. 28–60, Feb. 2004, doi: 10.1080/16501960410023732.

[2] L. K. Lindquist, H. C. Love, and E. B. Elbogen, "Traumatic Brain Injury in Iraq and Afghanistan Veterans: New Results From a National Random Sample Study," *J. Neuropsychiatry Clin. Neurosci.*, vol. 29, no. 3, pp. 254–259, Jul. 2017, doi: 10.1176/appi.neuropsych.16050100.

[3] K. McInnes, C. L. Friesen, D. E. MacKenzie, D. A. Westwood, and S. G. Boe, "Mild Traumatic Brain Injury (mTBI) and chronic cognitive impairment: A scoping review," *PLoS One*, vol. 12, no. 4, 2017, doi: 10.1371/journal.pone.0174847.

[4] P. Dall'Acqua et al., "Prefrontal Cortical Thickening after Mild Traumatic Brain Injury: A One-Year Magnetic Resonance Imaging Study," *J. Neurotrauma*, vol. 34, no. 23, pp. 3270–3279, 2017, doi: 10.1089/neu.2017.5124.

[5] G. I. Guberman, J. C. Houde, A. Ptito, I. Gagnon, and M. Descoteaux, "Structural abnormalities in thalamo-prefrontal tracks revealed by high angular resolution diffusion imaging predict working memory scores in concussed children," *Brain Struct. Funct.*, vol. 225, no. 1, pp. 441–459, 2020, doi: 10.1007/s00429-019-02002-8.

[6] P. Dall'Acqua et al., "Connectomic and surface-based morphometric correlates of acute mild traumatic brain injury," *Front. Hum. Neurosci.*, vol. 10, no. MAR2016, pp. 1–15, 2016, doi: 10.3389/fnhum.2016.00127.

[7] C. I. Kerley et al., "MRI correlates of chronic symptoms in mild traumatic brain injury," in *Medical Imaging 2020: Image Processing*, Mar. 2020, vol. 11313, p. 97, doi: 10.1117/12.2549493.

[8] J. L. R. Andersson and S. N. Sotiropoulos, "An integrated approach to correction for off-resonance effects and subject movement in diffusion MR imaging," *Neuroimage*, vol. 125, pp. 1063–1078, 2016, doi: 10.1016/j.neuroimage.2015.10.019.

[9] K. D. Cicerone and K. Kalmar, "Persistent postconcussion syndrome: The structure of subjective complaints after mild traumatic brain injury," *J. Head Trauma Rehabil.*, vol. 10, no. 3, pp. 1–17, Jun. 1995, doi: 10.1097/00001199-199510030-00002.

[10] A. J. Asman, A. S. Dagley, and B. A. Landman, "Statistical label fusion with hierarchical performance models.," *Proc. SPIE--the Int. Soc. Opt. Eng.*, vol. 9034, p. 90341E, Mar. 2014, doi: 10.1117/12.2043182.

[11] Y. Huo et al., "3D whole brain segmentation using spatially localized atlas network tiles," *Neuroimage*, vol. 194, pp. 105–119, Jul. 2019, doi: 10.1016/j.neuroimage.2019.03.041.

[12] I. Lyu, H. Kang, N. D. Woodward, M. A. Styner, and B. A. Landman, "Hierarchical spherical deformation for cortical surface registration," *Med. Image Anal.*, vol. 57, pp. 72–88, 2019, doi: 10.1016/j.media.2019.06.013.

[13] P. Parvathaneni et al., "Cortical Surface Parcellation Using Spherical Convolutional Neural Networks," in

*Lecture Notes in Computer Science (including subseries Lecture Notes in Artificial Intelligence and Lecture Notes in Bioinformatics)*, Oct. 2019, vol. 11766 LNCS, pp. 501–509, doi: 10.1007/978-3-030-32248-9_56.

[14] J. J. Koenderink and A. J. Van Doorn, "Surface shape and curvature scales," *Image Vis. Comput.*, vol. 10, no. 8, pp. 557–564, 1992.

[15] I. Lyu, H. Kang, N. D. Woodward, and B. A. Landman, "Sulcal depth-based cortical shape analysis in normal healthy control and schizophrenia groups," p. 1, 2018, doi: 10.1117/12.2293275.

[16] S. H. Kim *et al.*, "Development of cortical shape in the human brain from 6 to 24months of age via a novel measure of shape complexity," *Neuroimage*, vol. 135, pp. 163–176, Jul. 2016, doi: 10.1016/j.neuroimage.2016.04.053.

[17] I. Lyu, S. H. Kim, J. B. Girault, J. H. Gilmore, and M. A. Styner, "A cortical shape-adaptive approach to local gyrification index," *Med. Image Anal.*, vol. 48, pp. 244–258, Aug. 2018, doi: 10.1016/j.media.2018.06.009.

[18] I. Lyu, S. H. Kim, N. D. Woodward, M. A. Styner, and B. A. Landman, "TRACE: A Topological Graph Representation for Automatic Sulcal Curve Extraction," *IEEE Trans. Med. Imaging*, vol. 37, no. 7, pp. 1653–1663, Jul. 2018, doi: 10.1109/TMI.2017.2787589.

[19] J. D. Tournier, F. Calamante, and A. Connelly, "Determination of the appropriate b value and number of gradient directions for high-angular-resolution diffusion-weighted imaging," *NMR Biomed.*, vol. 26, no. 12, pp. 1775–1786, Dec. 2013, doi: 10.1002/nbm.3017.

[20] J.-D. Tournier, F. Calamante, and A. Connelly, "Improved probabilistic streamlines tractography by 2nd order integration over fibre orientation distributions," in *Proceedings of the International Society for Magnetic Resonance in Medicine*, 2010, p. 1670.

[21] J.-D. Tournier *et al.*, "MRtrix3: A fast, flexible and open software framework for medical image processing and visualisation," *bioRxiv*, p. 551739, 2019, doi: 10.1101/551739.

[22] Q. V. Le, A. Karpenko, J. Ngiam, and A. Y. Ng, "ICA with reconstruction cost for efficient overcomplete feature learning," *Adv. Neural Inf. Process. Syst. 24 25th Annu. Conf. Neural Inf. Process. Syst. 2011, NIPS 2011*, pp. 1–9, 2011.

[23] J. B. Kruskal, "Multidimensional scaling by optimizing goodness of fit to a nonmetric hypothesis," *Psychometrika*, vol. 29, no. 1, pp. 1–27, Mar. 1964, doi: 10.1007/BF02289565.

[24] M. R. Luo, "CIELAB," in *Encyclopedia of Color Science and Technology*, no. C, Berlin, Heidelberg: Springer Berlin Heidelberg, 2015, pp. 1–7.

[25] S. B. Eickhoff, K. Amunts, H. Mohlberg, and K. Zilles, "The Human Parietal Operculum. II. Stereotaxic Maps and Correlation with Functional Imaging Results," *Cereb. Cortex*, vol. 16, no. 2, pp. 268–279, Feb. 2006, doi: 10.1093/cercor/bhi106.

[26] B. D. McCandliss, L. Cohen, and S. Dehaene, "The visual word form area: expertise for reading in the fusiform gyrus," *Trends Cogn. Sci.*, vol. 7, no. 7, pp. 293–299, Jul. 2003, doi: 10.1016/S1364-6613(03)00134-7.

[27] I. A. J. van Kooten *et al.*, "Neurons in the fusiform gyrus are fewer and smaller in autism," *Brain*, vol. 131, no. 4, pp. 987–999, Apr. 2008, doi: 10.1093/brain/awn033.

[28] M.-E. Meadows, "Calcarine Cortex," in *Encyclopedia of Clinical Neuropsychology*, New York, NY: Springer New York, 2011, pp. 472–472.

[29] S. Japee, K. Holiday, M. D. Satyshur, I. Mukai, and L. G. Ungerleider, "A role of right middle frontal gyrus in reorienting of attention: a case study," *Front. Syst. Neurosci.*, vol. 9, Mar. 2015, doi: 10.3389/fnsys.2015.00023.

[30] S. Cutini *et al.*, "Selective activation of the superior frontal gyrus in task-switching: An event-related fNIRS study," *Neuroimage*, vol. 42, no. 2, pp. 945–955, Aug. 2008, doi: 10.1016/j.neuroimage.2008.05.013.

[31] M. L. Seghier, "The Angular Gyrus," *Neurosci.*, vol. 19, no. 1, pp. 43–61, Feb. 2013, doi: 10.1177/1073858412440596.

[32]  U. Hasson and P. Tremblay, "Neurobiology of Statistical Information Processing in the Auditory Domain," in *Neurobiology of Language*, Elsevier, 2016, pp. 527–537.